\documentclass[12pt]{article}

\usepackage{epsfig}
\usepackage{amsfonts}
\usepackage{amsopn}
\usepackage{amsmath}
\usepackage{verbatim,amsthm}

\title
{
\vskip -50 pt
\begin{flushright}
\normalsize\rm NORDITA-2010-66
\end{flushright}
\vskip 20 pt
On equivalence of the Komargodski-Seiberg action to the Volkov-Akulov action}

\author{
A. A. Zheltukhin $^{a,b,c}$\thanks{e-mail: aaz@physto.se}  \\ \\
$^a$ Kharkov Institute of Physics and Technology, \\
1, Akademicheskaya St., Kharkov, 61108, Ukraine \\  \\
$^b$ Fysikum, AlbaNova, Stockholm University, \\
106 91, Stockholm, Sweden \\ \\
$^c$ NORDITA,  \\
Roslagstullsbacken 23, 106 91 Stockholm, Sweden
}

\date{}

\begin{document}

\maketitle

\begin{abstract}

Equivalence between the Komargodski-Seiberg and the Volkov-Akulov supersymmetric  nonlinear Lagrangians of the Nambu-Goldstone fermions is proved in all orders in their interaction constant. Exact expression for the KS fermionic field through the VA fermion is found.

\end{abstract} 

\section{Introduction}

Recently Komargodski and Seiberg \cite{Seib} have developed  the $D=4 \  {\cal N}=1$ supercurrent approach to construct the low-energy effective Lagrangian of the  Nambu-Goldstone fermions \cite{VA}. The KS approach is based on the use of  connections between the linear and nonlinear realizations of supersymmetry and on the formalism of constrained superfields \cite{IK1}, \cite{R},  \cite{LR}. This  stimulates to deeper understanding of the NG fermions role in the minimal standard supersymmetric model, their couplings with other particles, astroparticle physics and the problem of  supersymmetry breaking (see e.g. \cite{ADGT}, \cite{AB}). 

 In the recent paper \cite{Zhep} we reported a difference between the KS Lagrangian \cite{Seib} and the well known Volkov-Akulov Lagrangian \cite{VA} which put forward the question about their equivalence. The difference is a consequence of the cancellation of 4-derivative terms in the VA Lagrangian earlier observed in  \cite{Kuz}.  To study this problem a simple alghorithmic procedure was developed in \cite{Zhep}, and the equivalence between the KS and VA Lagrangians was proved up to the first order in the interaction constant of the NG fermions.  

In the present paper we prove the equivalence in all orders in the interaction constant and derive an exact expression for the KS fermion through the VA fermion using this alghorithm.

\section{On a common  structure of the KS and the VA Lagrangians}

The algorithmic procedure \cite{Zhep} is based on the idea of the NG field redefinition earlier applied in  \cite{R}, and in  \cite{HK}, \cite{Kuz} while  studying nonlinear realizations of supersymmetric electrodynamics (see there additional refs.) However, practical realizations of the  procedure run against rather tedious calculations. It seems that the procedure \cite{Zhep} minimizes such type of  calculations, because of the special structure of bispinors in the polynomial expansion of the Majorana bispinor, representing the KS fermion, through the original VA fermion. The use of  this expansion allows to find an  exact expression for the KS fermion in terms of VA fermion. 

The Volkov-Akulov Lagrangian \cite{VA} (details may be found  in \cite{Z_Gra})
\begin{eqnarray} \label{VAlag} 
{\cal L}_{VA}= 
 \frac{1}{a} - \frac{i}{4}\bar\psi^{,m}\gamma_{m}\psi -
  \frac{a}{32}[(\bar\psi^{,m}\gamma_{m}\psi)^2{} -  \nonumber \\
(\bar\psi^{,n}\gamma_{m}\psi)(\bar\psi^{,m}\gamma_{n}\psi)
 ] + 
\frac{ a^2}{3!}\sum_{p} (-)^{p} T_{m}^{m} T_{n}^{n} T_{l}^{l},
\end{eqnarray}
where the $\sum_{p}$ corresponds to the sum in all permutations of the 
subindices in the products of the tensors $T^{m}_{n}$
\begin{eqnarray} \label{acterm}
T^{m}_{n}= -\frac{i}{4}\bar\psi^{,m}\gamma_{n}\psi,   \,\ \   \bar\psi^{,m}:=\partial^{m}\bar\psi\ .
\end{eqnarray} 
 The KS Lagrangian  \cite{Seib},  presented in the  bispinor Majorana representation up to the total derivative term, has the form 
\begin{eqnarray}\label{KSlag} 
{\cal L}_{KS}=
 \frac{1}{a} - \frac{i}{4}{\bar g}^{,m}\gamma_{m} g
- \frac{a}{16}[({\bar g}^{,m}g)^{2} + ({\bar g}^{,m}\gamma_{5}g)^{2}]  \nonumber \\
- (\frac{a}{16})^3[({\bar g}g)^{2} + ({\bar g}\gamma_{5}g)^{2}]
[
({\bar g}^{,m}g_{,m})^2 +
({\bar g}^{,m}\gamma_{5}g_{,m})^{2}],
\end{eqnarray}
where $g:= \sqrt{2}G$ and $a:=-1/f^2$, and the algebraic agreements, relations connecting bilinear covariants in the Weyl and the Majorana representations are similar to those used in \cite{Z_Gra}.
 One can see that each term in the VA and the KS Lagrangians  contains  the fermions and their derivatives only in the form of the  powers
 $(\partial\bar\psi\psi)^{n}$  and $(\partial{\bar g}  g)^{n}$,  respectively. These  combinations have various tensorial structures, but their dimensions are equal to $L^{-4n}$, and are inverse to the dimensions of $a^n$. This observation hints to seek for the expression of $g_a$ through $\psi_a$ just in the degrees $\partial\bar\psi\psi$.
 The conjecture results in the general polynomial \cite{Zhep} in the VA interaction constant $a$ 
\begin{eqnarray}\label{redef}
g= \psi + a\chi +  a^2\chi_{2} + a^3\chi_{3}.
\end{eqnarray}
The maximal degree of the polynomial (\ref{redef}) is less than four.  Actually, since  the products $ a^n\chi_{n}$ must have the same dimension as $\psi$, the Grassmannian Majorana  bispinors  $\chi, \, \chi_{2} , \, \chi_{3}$ may be constructed as $\psi$  multiplied by the above mentioned powers $(\partial\bar\psi\psi)^{n}$, i.e. they have the form $\chi_{n}\sim\psi(\partial\bar\psi\psi)^{n}$.
 These monomials are nilpotent and their maximal degree  $n=3$, because $\psi(\partial\bar\psi\psi)^{3}$ contains the maximal power of the  Grassmannian bispinor $\psi$ equal to four in the discussed case $D=4 \  {\cal N}=1$.
 The conjecture on the structure of the $\chi$-bispinors (\ref{redef})  opens a straightforward way to the construction of the redefined KS fermion $g$ through the VA fermion $\psi$, as it has been  manifested in \cite{Zhep}. Below we present the proof of the equivalence.

\section{Exact equivalence between the KS and the VA Lagrangians}

 Let us start the outlined  proof of the equivalence between the KS and the VA Lagrangians.  The substitution of the expansion (\ref{redef}) in the KS Lagrangian (\ref{KSlag}) and putting it equal to the VA Lagrangian  (\ref{VAlag}) yields the equations which  define the bispinors 
 $\chi, \, \chi_{2}$ and  $\chi_{3}$. Thus, the proof of the equivalence of the Lagrangians is reduced to the solutions of these equations.
The comparison of the terms of the same degree with respect to the constant $a$ in the redefined KS and the original VA Lagrangians, provides the algorithmic  way \cite{Zhep} to generate these equations.
 For instance, we have observed  that the spinors $\chi_{2}$ and  $\chi_{3}$ do not  contribute to the terms linear in $a$ in the redefined $L_{KS}$, and we  obtained the  equation defining the spinor $\chi$.  Then we are ready to consider the quadratic terms generating the  equation for $\chi_{2}$ and so on. 
Below we consider the procedure in more detail. 

The substitution of (\ref{redef}) into (\ref{KSlag}) 
and omission  of the total derivative terms yield the redefined KS kinetic term ${\cal K}_{0}$
\begin{eqnarray} \label{kinet}
{\cal K}_{0}:=  - \frac{i}{4}{\bar g}^{,m}\gamma_{m} g =  - \frac{i}{4}\bar\psi^{,m}\gamma_{m}\psi - \frac{i}{2}a(\bar\psi^{,m}\gamma_{m}\chi) -  \nonumber \\
\frac{i}{2}a^{2}[(\bar\psi^{,m}\gamma_{m}\chi_{2}) + \frac{1}{2}(\bar\chi^{,m}\gamma_{m}\chi)] - 
\frac{i}{2}a^{3}[(\bar\psi^{,m}\gamma_{m}\chi_{3}) + (\bar\chi^{,m}\gamma_{m}\chi_{2})]. 
\end{eqnarray}
The next term ${\cal K}_{1}$ from (\ref{KSlag}), linear in $a$ and quartic in fermionic fields, takes the form  
\begin{eqnarray} \label{1st}
{\cal K}_{1}: =- \frac{1}{16}a[(\bar g^{,m} g)^{2} + (\bar g^{,m}\gamma_{5}g)^{2}] = - \frac{1}{16}a[(\bar\psi^{,m} \psi)^{2} + (\bar \psi^{,m}\gamma_{5}\psi)^{2}] - \\
\frac{1}{16}a^{3}[(\bar\psi^{,m}\psi)(\bar\chi_{,m}\chi) + (\bar\psi^{,m}\gamma_{5}\psi)(\bar\chi^{,m}\gamma_{5}\chi)] \nonumber 
\end{eqnarray}
without  the  terms  quadratic in $a$, because of the relation
\begin{eqnarray} \label{ident}
(\bar g^{,m} g)=(\bar\psi^{,m}\psi) + a^{2}(\bar\chi^{,m}\chi) \,
\end{eqnarray}
and exclusion of the total derivate terms which arise from the relations 
$\bar\psi^{,m}\chi_{N} + \bar\chi^{,m}_{N}\psi=\partial^{m}(\bar\chi_{N}\psi)$.
Thus, the redefined term ${\cal K}_{1}$ generates  only linear and cubic in $a$ contributions to the VA Lagrangian. 
 The cubic in $a$ term ${\cal K}_{3}$ gives a non-zero contribution only for zero term from the expansion (\ref{redef}) $g=\psi$ 
\begin{eqnarray} \label{2nd}
{\cal K}_{3}= - \frac{1}{16}a^{3}
[({\bar\psi}\psi)^{2} + ({\bar\psi}\gamma_{5}\psi)^{2}]
[(\bar\psi^{,m}\psi_{,m})^2 + (\bar\psi^{,m}\gamma_{5}\psi_{,m})^2].  
\end{eqnarray}
This is due to the fact that the  contributions produced  by $\chi_{2}$ and $\chi_{3}$ contain  $\psi^5$ and higher identically vanishing factors constructed from  $\psi$ and  $\bar\psi$. 

Thus, we obtain the  redefined KS Lagrangian (\ref{KSlag}) presented  by the sum of  (\ref{kinet}),\, (\ref{1st}) and (\ref{2nd})
\begin{eqnarray}\label{KSredi}
{\cal L}_{KS}= \frac{1}{a} + {\cal K}_{0} + {\cal K}_{1} + {\cal K}_{3}.
\end{eqnarray}
Matching (\ref{KSredi}) with the VA Lagrangian (\ref{VAlag}) generates the  desired equations for the $\chi, \chi_{2}, \, \chi_{3}$ from (\ref{redef}). 

The equation for the spinor $\chi$, obtained  in \cite{Zhep},
\begin{eqnarray}\label{eqn1}
i(\bar\psi^{,m}\gamma_{m}\chi) = 
- \frac{1}{8}[({\bar\psi}^{,m}\psi)^{2} + ({\bar\psi}^{,m}\gamma_{5}\psi)^{2}] +
\nonumber  \\
 \frac{1}{16}[(\bar\psi^{,m}\gamma_{m}\psi)^2{} -
(\bar\psi^{,n}\gamma_{m}\psi)(\bar\psi^{,m}\gamma_{n}\psi)],
\end{eqnarray}
was simplified by cancellation of the common divisor $\bar\psi^{,m}$ to 
\begin{eqnarray}\label{eqn2}
\gamma_{m}\chi=  \frac{i}{8}[\psi({\bar\psi}_{,m}\psi) + \gamma_{5}\psi({\bar\psi}_{,m}\gamma_{5}\psi)] -
\nonumber  \\
 \frac{i}{16}[\gamma_{m}\psi (\bar\psi^{,n}\gamma_{n}\psi) -\gamma_{n}\psi
(\bar\psi^{,n}\gamma_{m}\psi)].
\end{eqnarray}
 Multiplication of Eq. (\ref{eqn2}) by $\gamma^{m}$ resulted in the general  solution
\begin{eqnarray}\label{solut1}
\chi= - \frac{i}{32}[(\gamma_{m}\psi)({\bar\psi}^{,m}\psi) + (\gamma_{m}\gamma_{5}\psi)({\bar\psi}^{,m}\gamma_{5}\psi)] -
\nonumber  \\
 \frac{i}{64}[3\psi (\bar\psi^{,m}\gamma_{m}\psi) + (\Sigma_{mn}\psi)
(\bar\psi^{,n}\gamma^{m}\psi)].
\end{eqnarray}
 We repeated here the derivation of the solution for $\chi$, because the same procedure will be applied to obtain general solutions of the equations for the spinors  $\chi_{2}$ and  $\chi_{3}$.

The equation for $\chi_{2}$ is derived by matching the   terms  quadratic in $a$  from (\ref{KSredi}) and  (\ref{VAlag}) 
 \begin{eqnarray}\label{eqn2}
(\bar\psi^{,m}\gamma_{m}\chi_{2}) = 
- \frac{1}{2}({\bar\chi}^{,l}\gamma_{l}\chi) + \frac{1}{3!2}\sum_{p} (-)^{p} (\bar\psi^{,m}\gamma_{m}\psi) T_{n}^{n} T_{l}^{l},
\end{eqnarray}
where the $\sum_{p}$ corresponds to the sum in all  the  permutations of the 
subindices in the products of the tensors $T^{n}_{n},\,T^{l}_{l}, \,(\bar\psi^{,m}\gamma_{m}\psi)$. This equation defines $\chi_{2}$ through the VA fermion $\psi$ and the solution (\ref{solut1}) for $\chi$.

The terms  cubic in $a$ from  (\ref{KSredi}) have to be mutually cancelled, because such terms are absent in the VA Lagrangian (\ref{VAlag}).
 It gives the third equation which defines the remaining spinor  $\chi_{2}$ 
\begin{eqnarray}\label{eqn3}
(\bar\psi^{,m}\gamma_{m}\chi_{3}) = 
- ({\bar\chi}^{,l}\gamma_{l}\chi_{2}) + \frac{i}{4}[(\bar\psi^{,m}\psi)(\bar\chi_{,m}\chi) +(\bar\psi^{,m}\gamma_{5}\psi)(\bar\chi_{,m}\gamma_{5}\chi)] + \nonumber
 \\
 \frac{2i}{16^{3}}[(\bar\psi\psi)^{2} + (\bar\psi\gamma_{5}\psi)^{2}]
[(\bar\psi^{,m}\psi_{,m})^2 +
(\bar\psi^{,m}\gamma_{5}\psi_{,m})^{2}].
\end{eqnarray}
 This equation defines $\chi_{3}$ through the solution (\ref{solut1}) for $\chi$ and  unknown yet spinor  $\chi_{2}$.  
To solve Eq. (\ref{eqn2}) for $\chi_{2}$ we observe that all its terms, besides the term $({\bar\chi}^{,l}\gamma_{l}\chi)$, contain the common divisor  $\bar\psi^{,m}$. However, the divisor  $\bar\psi^{,m}$  is hidden in $({\bar\chi}^{,l}\gamma_{l}\chi)$  as it follows from  the solution (\ref{solut1}) for $\chi$. Then we factorize this divisor in (\ref{solut1})  and present $\chi$ in the form 
\begin{eqnarray}\label{fact2}
\chi_{a}= A_{lab}{\bar\psi}^{,\,lb},
\end{eqnarray}
where the spinor matrix $A_{lab}$, carrying a 4- vector index $l$,  has the form
\begin{eqnarray}\label{matr2}
 A_{lab}: = \frac{i}{32}
[(\gamma_{l}\psi)_{a}\psi_{b} + (\gamma_{l}\gamma_{5}\psi)_{a}(\gamma_{5}\psi)_{b}] + \nonumber
\\
 \frac{i}{64}[3\psi_{a}(\gamma_{l}\psi)_{b} + (\Sigma_{nm}\psi)_{a} 
(\gamma^{n}\psi)_{b} ].
\end{eqnarray}
The substitution of (\ref{fact2}) into Eq. (\ref{eqn2}) transforms it to the  factorized form 
\begin{eqnarray}\label{eqn2'}
(\bar\psi^{,m}\gamma_{m}\chi_{2}) =
 \frac{1}{2} {\bar\psi}^{,\,mb}
({\bar\chi}^{,l}\gamma_{l}A_{m})_{b} + \frac{1}{3!2}\sum_{p} (-)^{p} (\bar\psi^{,m}\gamma_{m}\psi) T_{n}^{n} T_{l}^{l}
\end{eqnarray}
analogous with (\ref{eqn1}). After cancellation of the divisor $\bar\psi^{,m}$ of this equation we obtain solution for the bispinor  $\chi_{2a}$ 
\begin{eqnarray}\label{solut2}
\chi_{2a}= -\frac{1}{8}
\gamma^{mb}_{a.}(\bar\chi^{,l}\gamma_{l}A_{m})_{b} -
 \frac{1}{3!8}\sum_{p} (-)^{p} (\gamma^{m}\gamma_{m}\psi)_{a}
T_{n}^{n} T_{l}^{l}.
\end{eqnarray}

The similar procedure may be repeated to solve Eq. (\ref{eqn3}).
To avoid further complications let us  present the term
 $({\bar\chi}^{,l}\gamma_{l}\chi_{2})$ in Eq. (\ref{eqn3}) as 
\begin{eqnarray}\label{toder2}
({\bar\chi}^{,l}\gamma_{l}\chi_{2})= ({\bar\chi}_{2}^{,l}\gamma_{l}\chi) + \partial^l({\bar\chi}\gamma_{l}\chi_{2})
\end{eqnarray}
and omit  the terms which have the form of total derivatives.
 This trick allows to rewrite the relation  (\ref{eqn3}) in the factorized form  
\begin{eqnarray}\label{eqn3'}
(\bar\psi^{,m}\gamma_{m}\chi_{3}) =
{\bar\psi}^{,\,mb}
({\bar\chi}_{2} ^{,l}\gamma_{l}A_{m})_{b} + \nonumber
 \\
\frac{i}{4}[(\bar\psi^{,m}\psi)(\bar\chi_{,m}\chi) +(\bar\psi^{,m}\gamma_{5}\psi)(\bar\chi_{,m}\gamma_{5}\chi)] + \nonumber
 \\
 \frac{2i}{16^3}[(\bar\psi\psi)^{2} + (\bar\psi\gamma_{5}\psi)^{2}]
[(\bar\psi^{,m}\psi_{,m})^2 +
(\bar\psi^{,m}\gamma_{5}\psi_{,m})^{2}].
\end{eqnarray}
Again, as in the  previous cases, one can cancel the same  divisor  $\bar\psi^{,m}$ and get the  general solution for the bispinor $\chi_{3}$
\begin{eqnarray}\label{solut3}
\chi_{3a} =-\frac{1}{4}\gamma_{a.}^{mb}
({\bar\chi}_{2} ^{,l}\gamma_{l}A_{m})_{b} - \nonumber \\
\frac{i}{16}[(\gamma_{m}\psi)_{a}(\bar\chi^{,m}\chi) 
+ (\gamma_{m}\gamma_{5}\psi)_{a}(\bar\chi^{,m}\gamma_{5}\chi)] -  \nonumber\\
 \frac{i}{2(16)^{3}}[(\bar\psi\psi)^{2} + (\bar\psi\gamma_{5}\psi)^{2}]
[(\gamma_{m}\psi^{,m})_{a}(\bar\psi^{,l}\psi_{,l}) + \nonumber\\
(\gamma_{m}\gamma_{5}\psi^{,m})_{a}(\bar\psi^{,l}\gamma_{5}\psi_{,l})].
\end{eqnarray}

The comparison of the solutions (\ref{solut2}) and  (\ref{solut3}) for the bispinors $\chi_{Ja}$ (with $J=2,3$) shows that each of them contains similar combinations $(\gamma^{m}\bar\chi_{J}^{,l}\gamma_{l}A_{m})_{a}$.
 Using the definition (\ref{matr2}) of the matrix $A^{m}_{ab}$  allows to present this composite bispinor $(\gamma^{m}\bar\chi_{J}^{,l}\gamma_{l}A_{m})_{a}$ 
in the explicit form  
\begin{eqnarray}\label{exrepA}
(\gamma^{m}\bar\chi_{J}^{,l}\gamma_{l}A_{m})_{a}=-\frac{i}{32}[(\bar\chi_{J}^{,m}\psi)(\gamma_{m}\psi)_{a} - (\bar\chi_{J}^{,k}\Sigma_{km}\psi)(\gamma^{m}\psi)_{a} +
\nonumber\\
(\bar\chi_{J}^{,m}\gamma_{5}\psi)(\gamma_{m}\gamma_{5}\psi)_{a} - 
(\bar\chi_{J}^{,k}\Sigma_{km}\gamma_{5}\psi)(\gamma^{m}\gamma_{5}\psi)_{a}]  -    
\nonumber\\
\frac{i}{64}[12(\bar\chi_{J}^{,m}\gamma_{m}\psi)\psi_{a} - (\bar\chi_{J}^{,m}\gamma_{m}\Sigma_{kl}\psi)(\Sigma^{lk}\psi)_{a}].
\end{eqnarray}

Thus, we find all sought-for Majorana  bispinors $\chi,\,\chi_2$,\,$\chi_3$ from the relation (\ref{redef}) which expresses the KS fermion through the VA fermion  
\begin{eqnarray}\label{final}
g_{a}:=\sqrt{2}G_{a}= \psi_{a} + a(A_{m}\bar\psi^{,m})_{a} - 
\nonumber \\
 \frac{a^{2}}{8}[\,(\gamma^{m}\bar\chi^{,l}\gamma_{l}A_{m})_{a} +
 \frac{1}{3!}\sum_{p} (-)^{p} (\gamma^{m}\gamma_{m}\psi)_{a}
T_{n}^{n} T_{l}^{l}\,] -
\nonumber \\
\frac{a^{3}}{4}[\,(\gamma^{m}\bar\chi_{2}^{,l}\gamma_{l}A_{m})_{a} +
\frac{i}{4}[(\gamma_{m}\psi)_{a}(\bar\chi^{,m}\chi) + (\gamma_{m}\gamma_{5}\psi)_{a}(\bar\chi^{,m}\gamma_{5}\chi)] +
\nonumber \\
 \frac{2i}{(16)^{3}}[(\bar\psi\psi)^{2} + (\bar\psi\gamma_{5}\psi)^{2}]
[(\gamma_{m}\psi^{,m})_{a}(\bar\psi^{,l}\psi_{,l}) +
 \nonumber\\
(\gamma_{m}\gamma_{5}\psi^{,m})_{a}(\bar\psi^{,l}\gamma_{5}\psi_{,l})]\,].
\end{eqnarray}
The solution (\ref{final}) is presented in a compactified form.
When substituted into (\ref{final}), the solutions (\ref{solut1}), (\ref{solut2}) for the Majorana bispinors $\chi, \chi_{2}$ together with 
the expressions for matrix  $T^{m}_{n}$ (\ref{acterm}),\, $A_{m}$ (\ref{matr2}) and for the composite bispinors (\ref{exrepA}) 
 yield the representation for the KS fermion $G$ only through the VA fermion $\psi$ and its derivatives.

\section{Conclusion}

The equivalence between the Komargodski-Seiberg and the Volkov-Akulov $D=4 \  {\cal N}=1$  supersymmetric  actions  for the Nambu-Goldstone fermions is proved.  The representation for the KS fermionic field in terms of the VA fermion is found. This representation has a rather complicated form, but some  simplifications  may be  achieved using the Fierz  rearrengements.

\noindent{\bf Acknowledgments}

I would like to thank Sergei Kuzenko for his interest to this investigation.
I am grateful to the Department of Physics of Stockholm University 
and Nordic Institute for Theoretical Physics Nordita for kind hospitality. 
This research was supported in part by Nordita.


\begin{thebibliography}{99}

\bibitem{Seib}
Z. Komargodski and N. Seiberg,
JHEP {\bf 0909} 066, (2009);
arXiv: 0907.2441v3 [hep-th].
\bibitem{VA}
D.V. Volkov and V.P. Akulov, JETP Letters {\bf 16}  478 (1972);
 Phys. Lett. {\bf B 46}, 109  (1973);  Theor. Math. Phys.  {\bf 18}, 28 (1974).
\bibitem{IK1}
E.A. Ivanov and A.A. Kapustnikov,
J. Phys.  {\bf A 11} 2375, (1978); \\
J. Phys.  {\bf G 8} 167, (1982).
\bibitem{R}
M. Rocek, Phys.Rev. Lett. {\bf  41} 451, (1978). 
\bibitem{LR}
U. Lindstrom and M. Rocek, Phys. Rev. {\bf D 19} 2300, (1979).
\bibitem{ADGT}
I. Antoniadis, E. Dudas , D.M. Ghilencea, P. Tziveloglou,
Non-linear MSSM, arXiv:1006.1662 [hep-ph].
\bibitem{AB}
I. Antoniadis, M. Buican,
 Goldstinos, Supercurrents and Metastable SUSY Breaking in N=2 Supersymmetric Gauge Theories, arXiv:1005.3012 [hep-th].
\bibitem{Zhep}
A.A. Zheltukhin,
On the cancellation of 4-derivative  terms in the Volkov-Akulov action;
 arXiv:1003.4143 [hep-th].
\bibitem{Kuz}
S.M. Kuzenko and S.A. McCarthy, JHEP {\bf 05}  012  (2005).
\bibitem{HK}
T. Hatanaka and S.V. Ketov,
Phys. Lett. {\bf B 58} 265, (2004);  arXiv: hep-th/0310152.
\bibitem{Z_Gra}
A.A. Zheltukhin,
Dmitrij Volkov, super-Poincare group and Grassmann variables,
 Ann. Phys. (Berlin) {\bf 19}, No. 3-5, 177 (2010);
 arXiv:0911.0550 [hep-th];  Mod. Phys. Lett. {\bf 21}, No. 28, 2117 (2006).






\end{thebibliography}
\end{document}